\documentclass[a4paper,11pt]{article}
\usepackage{pos}
\usepackage{capt-of}

\title{Isospin-breaking corrections to light leptonic decays in lattice QCD+QED at the physical point}
\ShortTitle{ }

\author[a,b]{Peter Boyle}
\author*[a]{Matteo Di Carlo}
\author[a]{Felix Erben}
\author[a]{Vera G\"{u}lpers }
\author[a]{Maxwell T. Hansen}
\author[a]{Tim Harris}
\author*[c,d]{Nils Hermansson-Truedsson}
\author[a]{Raoul Hodgson}
\author[e,f]{Andreas J\"{u}ttner}
\author[a]{Fionn \'{O} h\'{O}g\'{a}in}
\author*[a]{Antonin Portelli}
\author[a,g]{James Richings }
\author[a]{Andrew Z.~N.~Yong }

\affiliation[a]{School of Physics and Astronomy, 
	University of Edinburgh, Edinburgh EH9 3FD, United Kingdom}
\affiliation[b]{Physics Department, Brookhaven National Laboratory, Upton, NY, USA}
\affiliation[c]{Albert Einstein Center for Fundamental Physics, Institute for Theoretical Physics, Universität Bern, Sidlerstrasse 5, CH-3012 Bern}
\affiliation[d]{Department of Astronomy and Theoretical Physics, Lund University, S\"{o}lvegatan 14A, 223 62 Lund, Sweden}
\affiliation[e]{CERN, Theoretical Physics Department, Geneva, Switzerland}
\affiliation[f]{School of Physics and Astronomy, University of Southampton, SO17 1BJ, Southampton, United Kingdom}
\affiliation[g]{EPCC, University of Edinburgh, EH8 9BT, Edinburgh, United Kingdom}

\emailAdd{matteo.dicarlo@ed.ac.uk,nils.hermansson-truedsson@thep.lu.se,antonin.portelli@ed.ac.uk}

\abstract{We report on the physical-point RBC/UKQCD calculation of the leading isospin-breaking corrections to light-meson leptonic decays. This is highly relevant for future precision tests in the flavour physics sector, in particular the first-row unitarity of the Cabibbo-Kobayashi-Maskawa matrix containing the elements $V_{us}$ and $V_{ud}$. The simulations were performed using Domain-Wall fermions for $2+1$ flavours, and with isospin-breaking effects included perturbatively in the path integral through order $\alpha$ and $(m_u - m_d)/\Lambda _{\mathrm{QCD}}$.  We use QED$_{\mathrm{L}}$ for the inclusion of electromagnetism, and discuss here the non-locality of this prescription which has significant impact on the infinite-volume extrapolation. }

\FullConference{%
The 39th International Symposium on Lattice Field Theory,\\
8th-13th August, 2022,\\
Rheinische Friedrich-Wilhelms-Universität Bonn, Bonn, Germany
}

\begin{document}
\maketitle

\section{Introduction}
\vspace{-0.4cm}
Leptonic decays of pions and kaons are important for flavour physics precision tests of the Standard Model (SM), since they give access to the ratio $|V_{us}|/|V_{ud}|$. The Cabibbo-Kobayashi-Maskawa (CKM) matrix elements $V_{us}$ and $V_{ud}$ are fundamental parameters in the SM expected to satisfy a unitarity relation with also $V_{ub}$, namely $	|V_{ud}|^2 +|V_{us}|^2 + |V_{ub}|^2 = 1$~\cite{Workman:2022ynf}. 
With a precision goal at the (sub-)per cent level, isospin-breaking effects in the strong and electromagnetic sector have to be included in theoretical predictions from e.g.~lattice field theory~\cite{Aoki:2021kgd}.

In the following we consider leptonic decays $P^+\rightarrow \ell ^+ \nu_{\ell}$, where $P^+$ is a pion or kaon, and the lepton $\ell^+$ a muon, including isospin-breaking effects in the strong sector ($m_u\neq m_d$) and from quantum electrodynamics (QED). To handle the problematic photon zero-momentum modes prohibiting QED to be straightforwardly defined in a finite volume with periodic boundary conditions, we employ the QED$_{\mathrm{L}}$ prescription~\cite{Hayakawa:2008an} where the spatial zero-momentum modes are dropped. Due to the long-range nature of QED, only the inclusive decay rate $\Gamma \left[ P^+ \rightarrow \ell ^+ \nu_\ell (\gamma)\right] = \Gamma _0\left[ P^+ \rightarrow \ell ^+ \nu_\ell \right] + \Gamma _1\left[ P^+ \rightarrow \ell ^+ \nu_\ell \gamma\right] $ yields an infrared (IR) finite result~\cite{Bloch:1937pw}. As originally laid out in Ref.~\cite{Carrasco:2015xwa}, one may separately study the virtual and real radiative corrected decay rates, $\Gamma _0$ and $\Gamma _{1}$, respectively,  by adding and subtracting a perturbatively calculated quantity with the same IR divergence. For instance, it is therefore possible to evaluate the virtual decay rate on the lattice in a spacetime of volume $T\times L^3$, and the real radiative decay perturbatively with a photon of mass $m_{\gamma}$ and energy less than some cut-off $\omega _\gamma$, so that
\begin{align}\label{eq:rm123ssep}
\Gamma \left[ P^+ \rightarrow \ell ^+ \nu_\ell (\gamma) \right] 	= 
\lim _{L\rightarrow \infty} \left\{ \Gamma _0 (L)- \Gamma _0 ^{\textrm{uni}}(L)  \right\} 
+ \lim _{m_\gamma \rightarrow 0} \left\{ \Gamma _0 ^{\textrm{uni}}(m_\gamma ) + \Gamma _{1}(\omega _{\gamma},m_{\gamma}) \right\} \, .
\end{align}
Here each term in brackets is IR-finite, with $L$ and $m_\gamma$ being used as IR regulators, and $ \Gamma _0 ^{\textrm{uni}}$ is a perturbatively calculated virtual decay rate independent of the internal meson structure (hence the superscript universal)~\cite{Carrasco:2015xwa,Lubicz:2016xro}. The quantity $ \Gamma _0 ^{\textrm{uni}}(L)$ was calculated for the first time in Refs.~\cite{Lubicz:2016xro,Tantalo:2016vxk}.
As discussed in Section~\ref{sec:fve},  $ \Gamma _0 ^{\textrm{uni}}$ subtracts the logarithmic IR-divergence and finite-volume effects through order $1/L$.  In Ref.~\cite{DiCarlo:2021apt}, an extension to a function $ \Gamma _0 ^{(n)}(L)$ subtracting structure-dependent higher-order finite-volume effects was proposed.  Here we use $ \Gamma _0 ^{(2)}(L)$ and therefore study the quantity
\begin{align}\label{eq:virtualdiff}
	\Gamma \left[ P^+ \rightarrow \ell ^+ \nu_\ell (\gamma) \right] 	& = 
	\lim _{L\rightarrow \infty} \left\{ 	\Gamma _0 (L) - \Gamma _0 ^{(2)}(L)   \right\} 
	+ \lim _{m_\gamma \rightarrow 0} \left\{ \Gamma _0 ^{\textrm{uni}}(m_\gamma ) + \Gamma _{1}(\omega _{\gamma},m_{\gamma}) \right\} 
	\nonumber
\\
& \equiv  
	\lim _{L\rightarrow \infty} \Gamma _P^{\mathrm{latt}} (L)
+ \lim _{m_\gamma \rightarrow 0}  \Gamma _P^{\mathrm{pert}} (m_\gamma)
\, ,
\end{align}
with $	\Gamma _0 (L)$ evaluated non-perturbatively and at leading order in $\alpha$ and $m_u-m_d$ on the lattice. 

The leading isospin-breaking corrections to the decay rates are defined through $ \Gamma _P^{\mathrm{latt}} (L) = \Gamma _{\mathrm{P}}^{\textrm{tree}} \left[ 1+\delta R_{P}^{\textrm{latt}}(L) - \delta R_{P}^{\textrm{(2)}}(L) \right] $ and $ \Gamma _P^{\mathrm{pert}} (m_\gamma) = \Gamma _{\mathrm{P}}^{\textrm{tree}} \, \delta R_P^{\textrm{pert}} (\omega _\gamma , m_\gamma )$ where the tree level decay rate is given by~\cite{Workman:2022ynf}
\begin{align}
	\Gamma ^{\textrm{tree}}_P = \frac{G_F^2}{8\pi}\, |V_{q_1 q_2}|^2\, m_\ell^2 \left( 1-\frac{m_\ell^2}{m_P^2}\right) ^2 \, m_P \, {f_{P}}^2 \, .
\end{align}
Here, $G_{F}$ is the Fermi constant, $V_{q_1 q_2}$ is the CKM element associated to the decaying meson $P$ comprised of valence quarks $q_{1,2}$, and ${f_{P}}$ is the isospin-symmetric decay constant of $P$. Introducing the combination $\delta R_P =  \delta R_{P}^{\textrm{latt}}(L) - \delta R_{P}^{\textrm{(2)}}(L) +  \delta R_P^{\textrm{pert}} (\omega _\gamma , m_\gamma )$, the ratio $|V_{us}|^2/|V_{ud}|^2$ can be obtained from the ratio of inclusive kaon and pion decay rates according to
\begin{align}
	\frac{|V_{us}|^2}{|V_{ud}|^2}
	 = 
	\frac{\Gamma \left[ K^+ \rightarrow \mu ^+ \nu_\mu (\gamma)\right] }{\Gamma  \left[ \pi^+ \rightarrow \mu^+ \nu_\mu (\gamma)\right]} 
	\left( \frac{m_{\pi}^2-m_{\mu}^2}{m_{K}^2-m_{\mu }^2}\right) ^2 \frac{m_{K}^3}{m_{\pi }^3 } \,  \left( \frac{{f_{\pi}}}{{f_{K}}}\right) ^2 \frac{1}{1+\delta R_{K}-\delta R_{\pi}}  \, . 
\end{align}
The ratio of decay constants as well as  $\delta R_{K\pi}=\delta R_{K}-\delta R_{\pi}$  can be predicted from the lattice~\cite{Aoki:2021kgd}, with the remaining parts taken from experiments. In the following we will focus on our calculation of $\delta R_{K\pi}$. For further details, we refer the reader to our paper Ref.~\cite{Boyle:2022lsi}. So far, there is only one other lattice prediction for this quantity, by the RM123/Southampton (RM123S) collaboration~\cite{Giusti:2017dwk,DiCarlo:2019thl}. 

\section{Calculating $\delta R_{K\pi}$ on the lattice}\label{sec:lattmeth}
\vspace{-0.4cm}
We consider the leptonic decay $P^+\rightarrow \ell^+ \nu _{\ell}$, with associated Euclidean 4-momenta in the rest frame of $P^+$, $p=(im_P,\mathbf{0} )$, $p_{\ell}=(i\omega _{\ell},\mathbf{p}_\ell)$ and $p_{\nu} = (i\omega _{\nu},-\mathbf{p}_\ell)$. We further denote polarisations of the final state leptons by $r$ and $s$, which have to be summed over to get the non-polarised decay rate. The finite-volume decay rate in the full QCD+QED theory can thus be written $\Gamma _{0}(L) = \mathcal{K} \, \left| \mathcal{M}\right| ^2$
where the kinematical factor $\mathcal{K}$ and the QCD+QED squared matrix element $\left| \mathcal{M} \right| ^2=\sum_{r,s} \left| \mathcal{M}^{rs}\right| ^2 $ are given by
\begin{align}
\mathcal{K} = \frac{G_F^2}{16\pi} \, \left| V_{q_1 q_2} \right| ^2 \, \frac{1}{2m_P} \left( 1-\frac{m_\ell^2}{m_P^2}\right) \, ,
 \quad 
 \mathcal{M}^{rs} = 
 \mathcal{Z} \, \left\langle \ell^+, \, r ,\, \mathbf{p}_\ell ; \nu _{\ell} , \, s , \, \mathbf{p}_{\nu}\right| O_{W}\left| P^+,\, \mathbf{0}\right\rangle  \, .
\end{align}
Here we introduced the four-fermion operator $O_W$ entering the effective Hamiltonian responsible for the decay, and its renormalisation factor $\mathcal{Z}$. We do not enter into the details regarding the renormalisation procedure here, and refer the reader to Ref.~\cite{DiCarlo:2019thl}. 

Next define $\left| \mathcal{M} \right|  ^2 = \mathcal{Z}^2 \, 4 \, m_\ell^2 \left( 1-m_\ell^2/m_P^2 \right) \, \left| \mathcal{A}_P \right| ^2 $. Plugging this into $\Gamma _{0}(L) = \mathcal{K} \, \left| \mathcal{M}\right| ^2$ and expanding to leading order in isospin breaking effects yields
\begin{align}\label{eq:dRPlattdef}
	\delta R_{P}^{\textrm{latt}} = 2\left( \frac{\delta \mathcal{A}_P}{{\mathcal{A}}_{P}^{(0)}} +\frac{\delta \mathcal{Z}}{  \mathcal{Z}_{0}} -\frac{\delta m_P}{ {m_P^{(0)}}}\right) 
	 \, ,
\end{align}
where  $\delta \mathcal{A}_P$ are the isospin-breaking corrections in the decay amplitude $\mathcal{A}_P$, and similar definitions hold for $\delta m_P$ and $\delta Z_P$. The isospin-symmetric quantities are $ \mathcal{Z}_0 \, {\mathcal{A}}^{(0)}_{P} =\mathcal{Z}_0 \, { \left\langle 0 \right| A^0 \left| P\right\rangle }^{(0)}  = i\, {m_{P}}^{(0)}\,  {f_{P}}$, where $A^0 = \bar{q}_2 \gamma_0 \gamma _5 q_1$ is the zeroth component of the axial current. From the mass-independence of $ \mathcal{Z}$, it follows that the difference $\delta R_{K\pi}$ is given by 
\begin{align}\label{eq:dRKpidef}
	\delta R_{K\pi}
	& = 
	2\left( \frac{\delta \mathcal{A}_K}{ {\mathcal{A}^{(0)}}_{K}} -\frac{\delta m_K}{ m_K^{(0)}}\right) 
	-2\left( \frac{\delta \mathcal{A}_\pi}{ {\mathcal{A}^{(0)}}_{\pi}}  -\frac{\delta m_\pi}{ m_\pi^{(0)}}\right)
	\nonumber \\
	&
- \left[ \delta R_{K}^{\textrm{(2)}}(L) - \delta R_{\pi}^{\textrm{(2)}}(L) \right]+ \left[  \delta R_K ^{\textrm{pert}} (\omega _\gamma , m_\gamma ) - \delta R_ \pi ^{\textrm{pert}} (\omega _\gamma , m_\gamma ) \right] 
	 \, . 
\end{align}

\begin{figure}[t!]
	\centering
		\includegraphics[height=0.2\textheight]{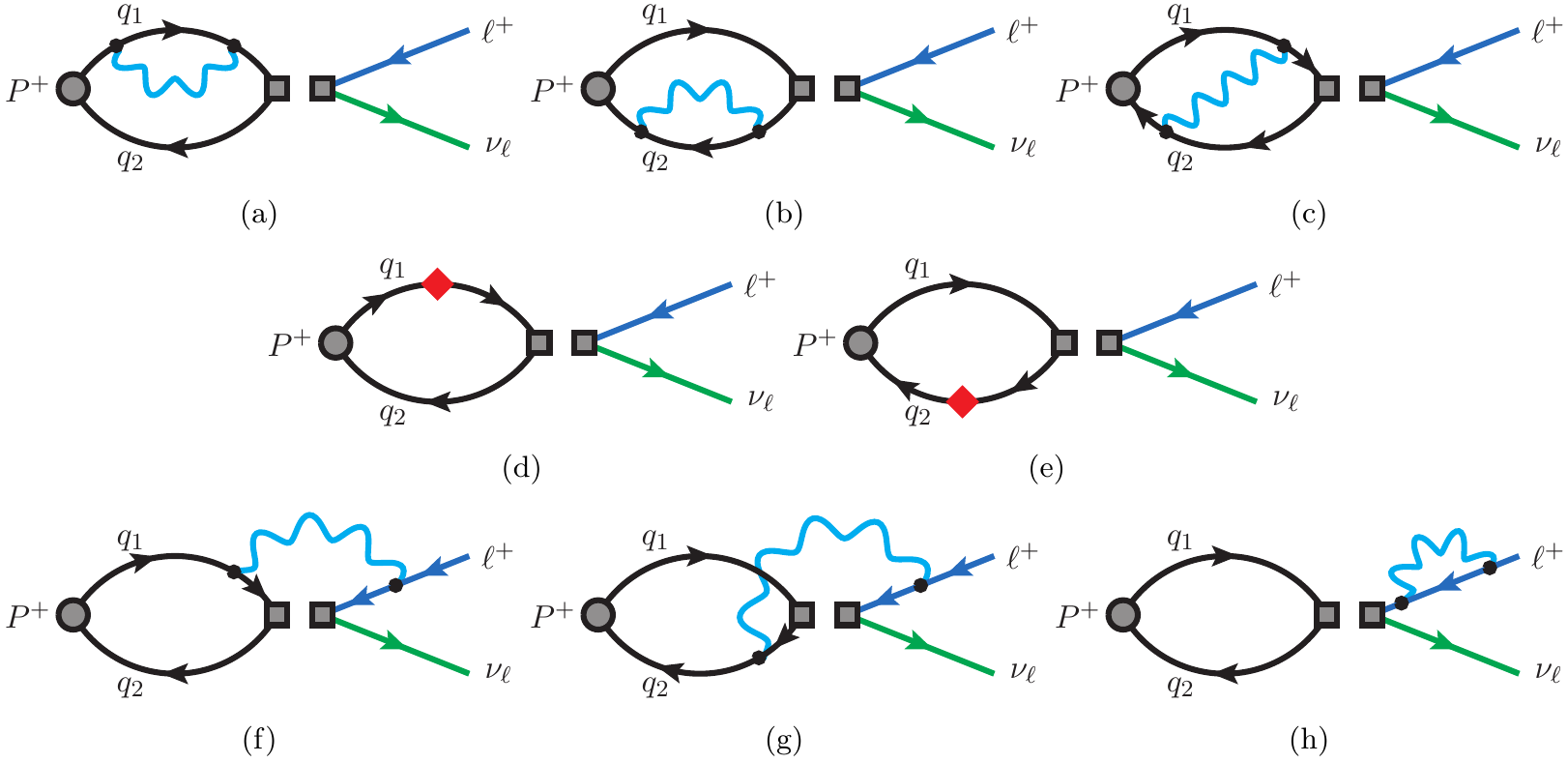}
	\caption{The leading connected isospin-breaking corrections to the decay $P^{+}\rightarrow \ell ^+ \nu _{\ell}$, where $P^+$ has valence quarks $q_1$ and $q_2$. The filled circles represent the meson interpolating operator, the squares the weak current and the diamonds scalar insertions. }	\label{fig:latticediagrams}
\end{figure}

We can get an understanding of the various isospin-breaking contributions $\delta \mathcal{A}_P$ by expanding the squared matrix element around the iso-symmetric point according to $	 \left| \mathcal{M} \right| ^2 =  \left| {\mathcal{M}}^{(0)} \right| ^2 + \delta _{\textrm{f}}  \left| \mathcal{M} \right| ^2 + \delta _{\textrm{nf}}  \left| \mathcal{M} \right| ^2$. 
The term $ \delta _{\textrm{f}}  \left| \mathcal{M} \right| ^2$ corresponds to the leading isospin-breaking corrections not involving the final-state lepton in the decay, which are diagrammatically depicted in diagrams (a)--(e) in Fig.~\ref{fig:latticediagrams}. As can be seen, the leptonic parts are factorised. Conversely, $ \delta _{\textrm{nf}}  \left| \mathcal{M} \right| ^2$ contains the non-factorisable diagrams (f)--(g). The remaining diagram, (h), does not need to be considered as it has purely perturbative nature and thus cancels in the difference in~(\ref{eq:virtualdiff}). Note that we employ an electroquenched approximation where sea quarks are neutral. These neglected disconnected diagrams are expected to be small~\cite{DiCarlo:2019thl}. 

\subsection{Extracting $\delta R_{K\pi}$ from correlators}
\vspace{-0.2cm}
The non-perturbative quantities needed for $\delta R_{P}^{\textrm{latt}}$ in~(\ref{eq:dRPlattdef}) can be extracted from Euclidean correlation functions on the lattice. We denote the finite time-extent by $T$, and here choose the meson to be created at time $-t$, the weak current to be at the origin and the lepton a time-separation $t_{\ell}$ away from the weak current. 
The factorisable contributions, which do not couple to the leptons, can be obtained by studying the two correlators
\begin{align}
	C_{\textrm{PP}}(t) & = \int d^3 \mathbf{x} \, \left\langle 0 \right| \mathbb{T} \left\{ A^0 (0,\mathbf{0})\,  \phi ^\dagger (-t,\mathbf{x}) \right\} \left| 0 \right\rangle \, ,
	  \\
	  C_{\textrm{PA}}(t) & = \int d^3 \mathbf{x} \, \left\langle 0 \right| \mathbb{T} \left\{ \phi (0,\mathbf{0}) \, \phi ^\dagger (-t,\mathbf{x}) \right\} \left| 0 \right\rangle \, ,
\end{align}
where $\phi$ is a pseudoscalar meson interpolator with the appropriate flavour structure.
 The above correlators are defined in the full theory, i.e.~QCD+QED, but we will expand them around the iso-symmetric point through leading order in isospin breaking. Performing spectral decompositions of the above correlators in the iso-symmetric limit, and retaining only the leading forward and backward propagating exponentials, yields
\begin{align}
	C_{\textrm{PP}}^{(0)}(t) &  = \frac{ |Z_{P}|^{2}}{2\, m_P^{(0)}} \, \left[ e^{-m_P^{(0)}t}+e^{-m_P^{(0)}\left(T-t\right)}\right] + \ldots \, ,
	\\
	C_{\textrm{PA}}^{(0)}(t) & =   \frac{ Z_P \, {\mathcal{A}^{(0)}}_{P}}{2\, m_P^{(0)}} \, \left[ e^{-m_P^{(0)}t}-e^{-m_P^{(0)}\left(T-t\right)}\right] + \ldots \, ,
\end{align}
where $Z_P = \langle P, \mathbf{0} | \phi ^{\dagger}(0)| 0 \rangle$ and the ellipses contain sub-leading exponentials. It is from the above spectral decompositions that the iso-symmetric quantities in~(\ref{eq:dRPlattdef}) can be fitted. Defining $\delta C_{\textrm{PA}}(t)$ as the leading isospin-breaking contribution to $	C_{\textrm{PA}}(t) $ gives the factorisable correction $\delta _{\textrm{f}}\mathcal{A}_P$ from the leading finite-$T$ behaviour of the ratio
\begin{align}
	\frac{\delta C_{\textrm{PA}}(t)}{C_{\textrm{PA}}^{(0)}(t)} & =  \frac{\delta _{\textrm{f}} \mathcal{A}_{P}}{ {\mathcal{A}^{(0)}}_{P}}+\frac{\delta Z_P}{ Z_P} -\frac{\delta m_P}{m_P^{(0)}} \, f_{PA}(t) + \ldots \, ,
	\\
	f_{PA}(t) & = 1+ m_P^{(0)} \left\{ \frac{T}{2}-\left( t-\frac{T}{2}\right) \, \coth \left[ m_P^{(0)}\left( t-\frac{T}{2}\right)  \right] \right\} \, .
\end{align}
A similar procedure can be used for $C_{\textrm{PP}}(t)$. 

What remains is to determine the non-factorisable contribution $\delta _{\textrm{nf}}\mathcal{A}_{P}$. In a fashion similar to above, we define a correlation function $C_{P\ell}(t)$ containing the meson, the weak current as well as the leptonic part. This can be appropriately traced with spinors to yield a scalar object. However, the lepton propagator between the weak current at the origin and the external lepton at $t_\ell$ affects the finite-$T$ behaviour severely. By constructing projectors onto either the forward or backward propagating signal, the numerical analysis can be simplified. For brevity, we refrain from defining this correlator in detail, and simply show the leading finite-$T$ behaviour when projecting onto the forward propagating lepton, namely
\begin{align}
	\frac{\delta C_{P\ell} (t)}{\overline{C}_{P\ell }(t)}  = \frac{\delta _{\textrm{nf}}\mathcal{A}_{P}}{ {\mathcal{A}^{(0)}}_{P}} \, f_{P\ell}(t) + \ldots \, ,
	\quad
	f_{P\ell}(t) = \frac{1}{2}\left\{1+\kappa _\ell -(1-\kappa _\ell) \, \coth \left[ m_P^{(0)}\left(t-\frac{T}{2}\right)\right] \right\} \, ,
\end{align}
where $\kappa _\ell$ parametrises the backward propagating signal. 

We thus see that all the ingredients needed for $\delta R_{P}^{\textrm{latt}}$ in~(\ref{eq:dRPlattdef}) can be obtained from Euclidean correlation functions on the lattice. We do simultaneous correlated fits of the factorisable and non-factorisable correlators, respectively. We choose optimal fit ranges through a genetic algorithm for the factorisable case, and construct $\delta R_{K\pi}$ from these using AIC-based model averaging as in e.g.~Refs.~\cite{Borsanyi:2014jba,Jay:2020jkz}. This approach results in a distribution of possible $\delta R_{K\pi}$, from which we choose our prediction as the median and estimate the associated statistical and systematic uncertainties. 

\subsection{Simulation details}
\vspace{-0.2cm}
For the simulations needed to generate the correlators discussed above we make use of {\sc Grid}~\cite{Boyle:2016lbp} and {Hadrons}~\cite{antonin_portelli_2022_6382460}. We have temporal and spatial extents $T/a = 96$ and $L/a = 48$, respectively, and use domain-wall fermions close to the physical point for $L_{s}/a =24$ and $a M_5=1.8$~\cite{RBC:2014ntl}. We use 60 statistically independent QCD gauge configurations generated using the Iwasaki action~\cite{IWASAKI1985141} by the RBC/UKQCD collaboration. Our valence- and sea-quark masses are  $\hat{m}_u = \hat{m}_d = 0.00078$ and $\hat{m}_s = 0.0362$, where hat denotes lattice units. The ensemble pion mass is $m_\pi = 139.15(36) $ MeV, and the inverse lattice spacing $a^{-1} = 1.7295(38)$ GeV. We thus have a single lattice spacing and volume. 

The correlators are created using sequential propagators, with 96 Coulomb gauge-fixed wall sources per configuration. The muon momentum $\mathbf{p}_\ell$ fixed in the direction $(1,1,1)$ is injected via twisted boundary conditions, and the muon propagator is evaluated for 8 different source-sink separations. With the weak current at time $0$, these are $t_\ell \in \{12,16,20,24,28,32,36,40\}$. The photons are implemented in $\textrm{QED}_{\textrm{L}}$, with the photon fields sampled from a Gaussian distribution, and we use a renormalised local vector current. 

\subsection{Defining the iso-symmetric theory}\label{sec:schemes}
\vspace{-0.2cm}
The bare parameters of the full QCD+QED theory can be unambiguously determined by requiring that a complete set of experimental hadronic masses be reproduced. Once determined, any physical quantity can be predicted. Determining the isospin-breaking corrections to an observable, on the other hand, requires the definition of an iso-symmetric point. Consider a physical observable $X^\phi$.  This can be separated into three terms according to $	X^\phi = {X}^{(0)} + \delta X^\gamma + \delta X^{\textrm{SIB}} $, where $ {X}^{(0)}$ is the iso-symmetric quantity, $\delta X^\gamma$ contains the electromagnetic corrections where $\alpha \neq 0$, and $\delta X^{\textrm{SIB}}$ includes the strong isospin-breaking effects from $ m_u\neq m_d$ for $\alpha =0$. Such a separation, however, requires additional conditions and constitutes the choice of a scheme. 
Therefore, when comparing predictions of isospin-breaking corrections, scheme ambiguities have to be taken into account. This is particularly important since the ambiguities in principle can be of the same size as the predictions themselves. Relations between different schemes can be obtained due to the small numerical size of isospin-breaking effects, which implies that unphysical theories to a good approximation are within a linear deviation from the physical point. 

In our calculation we have 3 flavours. Denoting lattice units by a hat, we tune our parameters in the full theory to reproduce the ratios
\begin{align}
	\frac{\hat{m}_{P}}{\hat{M}_{\Omega ^{-}}} = \frac{m ^{\textrm{PDG}}_{P}}{M^{\textrm{PDG}}_{\Omega^-}} \, ,
\end{align}
where $P=\pi^+,K^+ , K^0$ and PDG denotes the Particle Data Group values from Ref.~\cite{Workman:2022ynf}.  In addition, we can define the lattice spacing as $a=(\hat{M}_{\Omega ^-})/M^{\textrm{PDG}}_{\Omega ^-}$. To define the iso-symmetric point we employ the same scheme as the BMW collaboration in Ref.~\cite{PhysRevLett.111.252001}. In short, this is done by considering non-physical, neutral, purely connected mesons with quark content $\bar{q}q$ and masses $M_{\bar{q}q}$, which are combined into the three variables (in the full theory)
\begin{align}
	M^2_{ud} = \frac{1}{2} \left( M^2_{\bar{u}u}+M^2_{\bar{d}d}\right) \, , \quad \Delta M^2 = M^2_{\bar{u}u}-M^2_{\bar{d}d} \, , \quad 2\, M^2_{K\chi } = M^2_{K^+} + M_{K^0}^2-M_{\pi^+}^2 \, .
\end{align}
The scheme is then defined by requiring that the meson masses above coincide in the full theory and QCD, as well as imposing the conditions $ (\hat{M}^{(0)}_{ud}/\hat{M}_{\Omega ^{-}} ) ^{2 } = (\hat{M}_{ud} / \hat{M}_{\Omega ^{-}})^2$ , $ (\hat{M}_{K\chi } ^{(0)}/\hat{M}_{\Omega ^{-}})^{2}  = (\hat{M}_{K\chi } /\hat{M}_{\Omega ^{-}})^{2}$ and $ (\Delta \hat{M}^{(0)}/\hat{M}_{\Omega ^{-}}) ^{2} = 0$ between the full and iso-symmetric theories.

\section{Finite-volume effects}\label{sec:fve}
\vspace{-0.4cm}
The finite-volume effects in the decay rate $	\Gamma _{0}^{(n)}(L)$ are contained in the correction $\delta R_{P}^{(n)}(L)$ defined through
\begin{align}
		\Gamma _{0}^{(n)}(L) 
	& =
	\Gamma _{P}^{\textrm{tree}} \left[ 1+ \delta R_{P}^{(n)}(L) \right]\, ,
	\\
\delta R_{P}^{(n)}(L)
	& = 2\, \frac{\alpha }{4\pi} \left[
	 \widetilde{Y}_P(L) + \sum_{j=0}^{n}\frac{1}{(m_PL)^j}\, Y_{P,\, j} 
	 \right]
	 \, .
\end{align}
The universal $\widetilde{Y}_P (L)$ contains the infinite-volume contribution as well as the logarithmic divergence in $L$. The numerical coefficients $Y_{P,\, j}$ in general depend on the structure of the decaying mesons. However, it was proven in Ref.~\cite{Lubicz:2016xro} that through order $1/L$ the coefficients are universal, i.e.~structure-independent, and they were determined within point-like scalar QED. 

The structure-dependent contributions at higher order can be determined using the approach in Ref.~\cite{DiCarlo:2021apt}. At present, the coefficients are known fully through order $1/L^2$ and can be found in Ref.~\cite{DiCarlo:2021apt}. The structure dependence in $Y_{P,\, 2}$ is encoded in the on-shell axial-vector form factor $F_A^P$ from the real radiative decay $P\rightarrow \ell \nu _\ell \gamma $, which is known experimentally~\cite{Workman:2022ynf}, from chiral perturbation theory~\cite{Cirigliano2012} and the lattice~\cite{Desiderio:2020oej}. However, it was shown in Ref.~\cite{DiCarlo:2021apt} that due to the smallness of $F_A^P$, the structure-dependence at order $1/L^2$ is only a per cent level contribution to $Y_{P,\, 2}$, implying that $Y_{P,\, 2}\approx Y_{P, \, 2}^{\textrm{pt}}$ is a good approximation. 

At order $1/L^3$, only the point-like approximation $Y_{P, \, 3}^{\textrm{pt}}$ of the full coefficient $Y_{P, \, 3}= Y_{P, \, 3}^{\textrm{pt}}+Y_{P,\, 3}^{\textrm{sd}}$ is currently known. The structure-dependent piece $Y_{P,\, 3}^{\textrm{sd}}$ has yet to be determined, which is complicated due to non-local effects arising at order $1/L^3$ in $\textrm{QED}_\textrm{L}$~\cite{DiCarlo:2021apt}. The point-like result is
\begin{align}\label{eq:y3point}
	Y_{P,\, 3}^{\textrm{pt}} = 		\frac{32\pi^2 c_{0} \, [2+(m_{\ell}/m_P)^2] }{ [ 1+(m_{\ell}/m_P)^2]^3 } \, .
\end{align}
Here $c_0 = -1$ is a finite-volume coefficient.
As a final remark on the finite-volume expansion, all the point-like contributions at order $1/L^4$ and higher vanish as is easily shown using the approach in Ref.~\cite{DiCarlo:2021apt}.

In our analysis we subtract $\delta R_{P}^{(2)}(L)$ from $\delta R_{P}^{\textrm{latt}}(L)$ to yield $\delta R_{K\pi}$ as in~(\ref{eq:dRKpidef}). At our simulated volume $L/a = 48$ we then use the point-like contribution $\delta R_{P}^{(3),\, \textrm{pt}}(L) -\delta R_{P}^{(2)}(L)$ as a systematic error. Denoting $\delta R_{K\pi}^{(n)}(L) = \delta R_{K}^{(n)}(L)  - \delta R_{\pi}^{(n)}(L) $ we find at the simulation point for $L_{48}=48a$  
\begin{align}
	\delta R^{(1)}_{K\pi} (L_{48}) \approx -0.00468\, , \quad 	\delta R^{(2)}_{K\pi} (L_{48}) \approx -0.00730  \, , 	\quad \delta R^{(3), \, \textrm{pt}}_{K\pi} (L_{48}) \approx -0.00337 \, .
\end{align}
The relative shift in going from $1/L^2$ to $1/L^3$ is roughly $-54\%$, which means that we have a large systematic error from not knowing the full coefficient $Y_{P,\, 3}$. We are currently studying possible ways to determine the unknown structure-dependent $Y_{P,\, 3}^{\textrm{sd}}$. 

Finally, we present the importance of analytical knowledge of the finite-volume dependence for the infinite-volume extrapolation. For this we use the published  $\delta R_{K\pi}$ data of Ref.~\cite{DiCarlo:2019thl}, which was kindly provided to us by the authors. The data was produced for non-physical pions and kaons of masses $m_\pi \approx 320 $ MeV and $m_{K}\approx 580$ MeV, respectively, and for several volumes. We here investigate the impact on the infinite-volume extrapolation of this data from the recent observation that the structure-dependence in $Y_{P,\, 2}$ is negligible, which was not known at the time of Ref.~\cite{DiCarlo:2019thl}.

\begin{figure}[t!]
	\centering
	\begin{minipage}{.5\textwidth}	
		\centering
		\includegraphics[height=0.25\textheight]{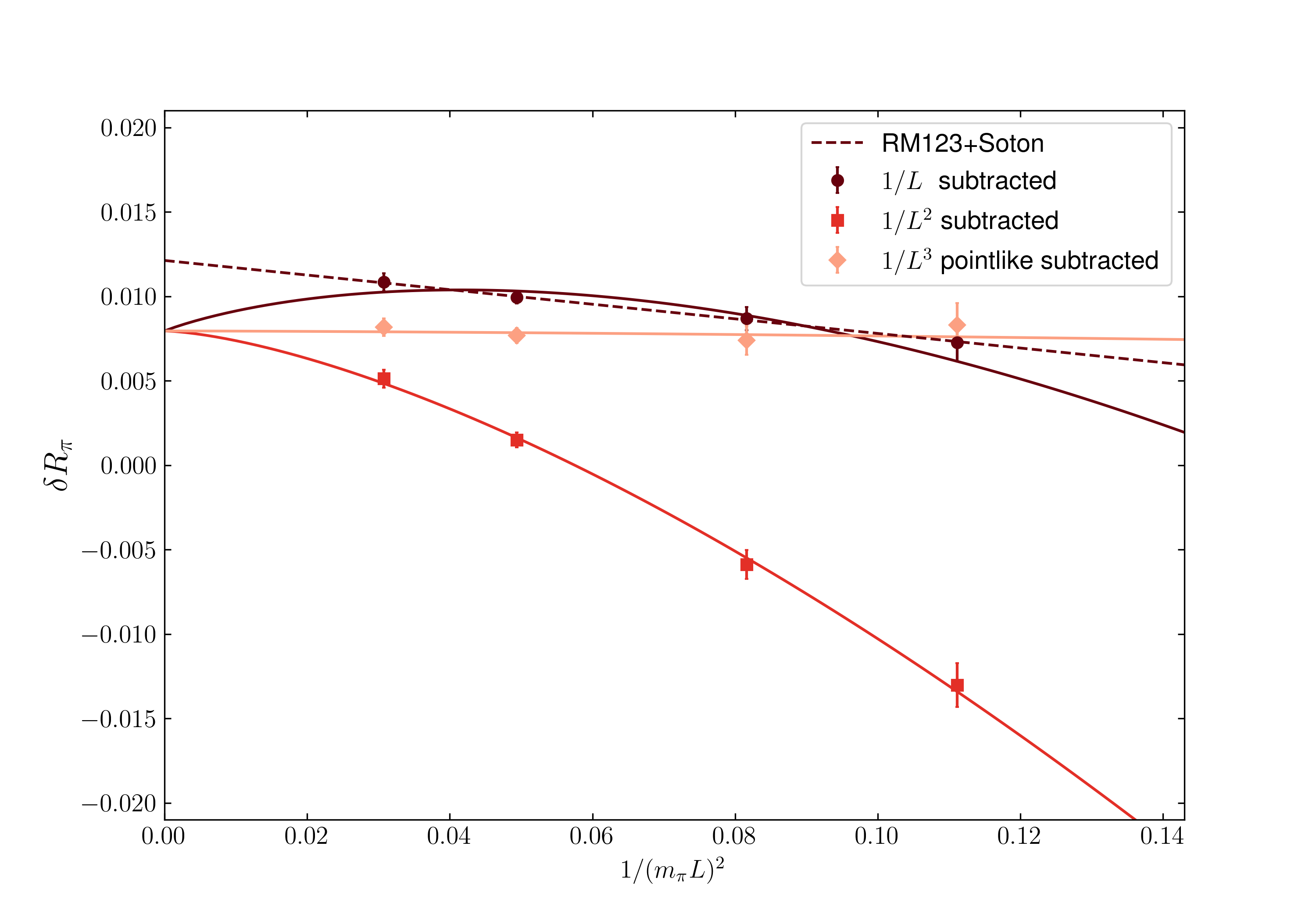}
		\caption*{(a)}
	\end{minipage}%
	\begin{minipage}{0.5\textwidth}	
		\centering
		\includegraphics[height=0.25\textheight]{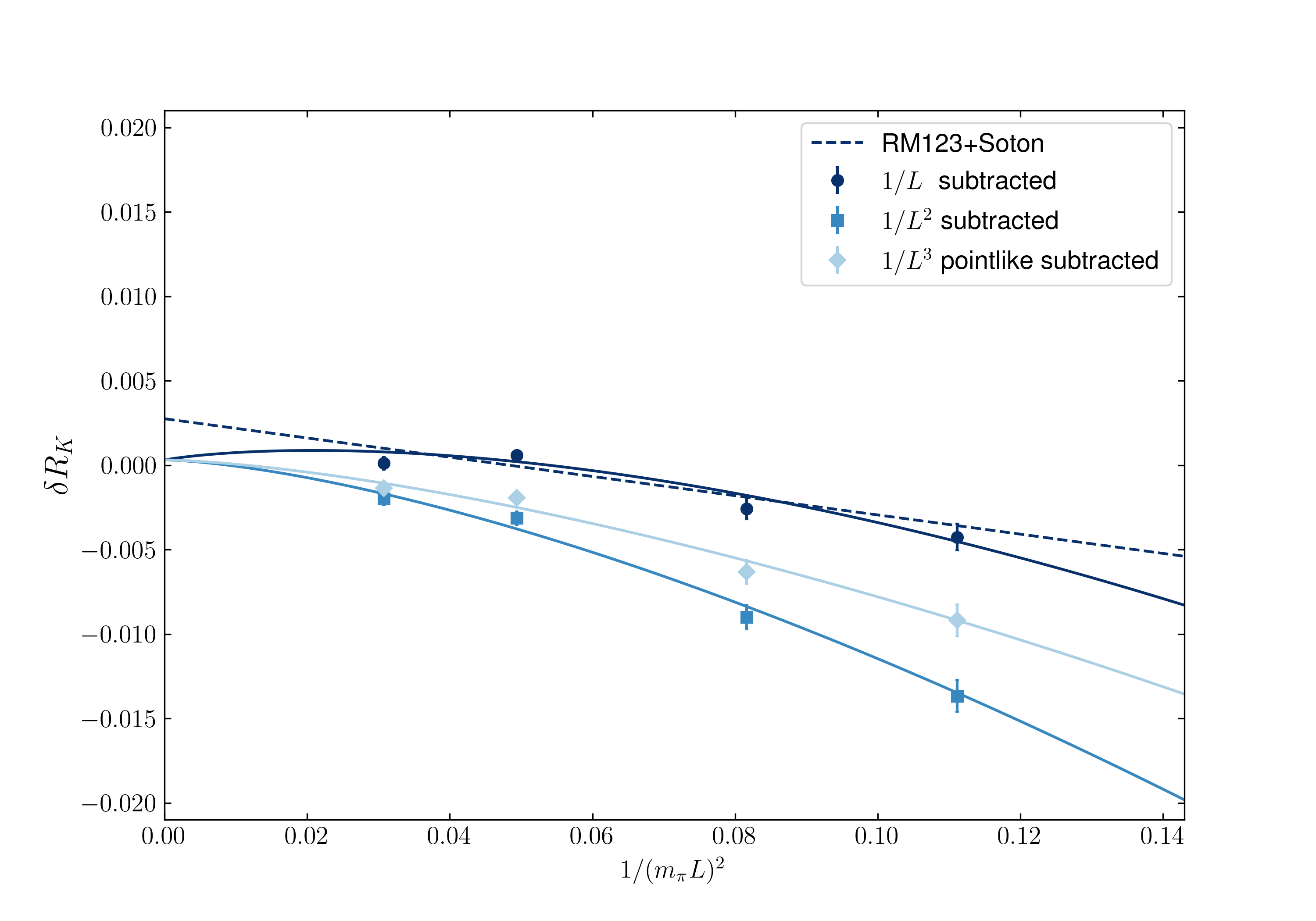}
		\caption*{(b)}
	\end{minipage}
	\caption{The volume dependence of $\delta R_{P}$ for  (a) pions and (b) kaons, subtracted with $\delta R_P^{(n)}(L)$ for different $n$. The lattice data points are from Ref.~\cite{DiCarlo:2019thl}. }\label{fig:rm123sdata}
\end{figure}

In Figs.~\ref{fig:rm123sdata}(a)--(b) the volume-dependence of the pion and kaon data for $\delta R_{P}$ is shown,  with different orders of $\delta R_{P}^{(n)}(L) $ subtracted. The circular points in the figures correspond to the results of Ref.~\cite{DiCarlo:2019thl} where the universal $\delta R_{P}^{(1)}(L) $ are subtracted (labelled $1/L$ subtracted). As can be seen, a linear extrapolation to the infinite-volume limit (dashed line) describes the data well. In Ref.~\cite{DiCarlo:2019thl} the point-like approximation of $\delta R_{P}^{(2)}(L) $ was then subtracted, which here corresponds to the square points, and due to the residual slope the authors of Ref.~\cite{DiCarlo:2019thl} concluded that there is significant structure-dependence at order $1/L^2$. However, from Ref.~\cite{DiCarlo:2021apt} we now know that the structure-dependence in $Y_{P,\, 2}$ is negligible, so that the slope of the $1/L^2$ subtracted data in Fig.~\ref{fig:rm123sdata} must be from the $1/L^3$ coefficient. 
We therefore also subtract the point-like $\delta R_{P}^{\textrm{pt},\, (3)}(L) $ to yield the diamond points, and perform a fit of the form $a+b/L^3$ to those points (solid line in the figure). The $1/L^3$ ansatz for the volume-dependence describes the data well and estimates the remaining $1/L^3$-dependence. Using the fitted curve and adding back the analytically known coefficients yields the remaining solid lines, which also give a good description of the data. This means that a sizeable $1/L^3$  dependence mimics a $1/L^2$ behaviour in the range of volumes in the figure.

Although the infinite-volume extrapolations for the dashed and the solid curves do not agree in the figure, we stress that this difference can be washed out in predictions of $\delta R_{K\pi}$. For instance, the numerical analysis to obtain the final value of $\delta R_{K\pi}$ in Ref.~\cite{DiCarlo:2019thl} involves several steps beyond the infinite-volume extrapolation, such as a simultaneous chiral extrapolation to physical meson masses, and the difference observed here might be well within the associated uncertainties. In conclusion, analytical knowledge of the finite-volume dependence is crucial when extrapolating to infinite volume, and there is clear need to determine the full coefficient $Y_{P,\, 3}$. 

\section{A prediction for $\delta R_{K\pi}$}\label{sec:prediction}
\vspace{-0.4cm}
We now present our prediction for the leading isospin-breaking to ratio of light leptonic decays,
\begin{align}
	\delta R_{K\pi} = -0.0086(3)_{\textrm{stat.}}\left(^{+11}_{-4}\right)_\textrm{fit} (5)_{\textrm{disc.}} (5)_{\textrm{quench.}} (39)_{\textrm{vol.}}  \, .
\end{align}
As can be seen the dominant contribution comes from the finite-volume uncertainty. Further details on the error budget is provided in our paper~\cite{Boyle:2022lsi}. This value for $\delta R_{K\pi} $ can be compared to those from RM123S~\cite{DiCarlo:2019thl} and chiral perturbation theory ($\chi$PT)~\cite{Cirigliano:2011tm}, which respectively are
\begin{align}
	\delta R_{K\pi}^{\textrm{RM123S}}  = -0.0126(14) \, ,
	\quad
	\delta R_{K\pi}^{\chi \textrm{PT}}  = -0.0112(21) \, . 
\end{align}
All three values are in agreement, although our error bars are very large due to the finite-volume systematics. It is clear that the finite-volume issue has to be resolved in the future. 

\section{Conclusions}\label{sec:summary}
\vspace{-0.4cm}
In these proceedings we have presented the RBC/UKQCD prediction for the leading isospin-breaking corrections to leptonic decays of pions and kaons, encoded in $\delta R_{K\pi}$.  Further details on our calculation are presented in Ref.~\cite{Boyle:2022lsi}. Knowledge of this quantity is essential for future flavour physics precision tests of the Standard Model, particularly as it is required for testing CKM-matrix unitarity at (sub-)per cent precision. 

\section*{Acknowledgements}
\vspace{-0.4cm}
We thank the RM123S collaboration for kindly sharing their lattice data for our study of the finite-volume dependence. This work used the DiRAC Extreme Scaling service at the University of Edinburgh, operated by the Edinburgh Parallel Computing Centre on behalf of the STFC DiRAC HPC Facility (\url{www.dirac.ac.uk}). This equipment was funded by BEIS capital funding via STFC capital grant ST/R00238X/1 and STFC DiRAC Operations grant ST/R001006/1. DiRAC is part of the National e-Infrastructure. 
P.B. has been supported in part by the U.S.~Department of Energy, Office of Science, Office of Nuclear Physics under the Contract No.~DE-SC-0012704 (BNL).
M.D.C., F.E., T.H., V.G., M.T.H., and A.P. are supported in part by UK STFC grant ST/P000630/1. Additionally M.T.H. is supported by UKRI Future Leader Fellowship MR/T019956/1.   
F.E., V.G., R.H., F.\'Oh., A.P. and A.Z.N.Y. received funding from the European Research Council (ERC) under the European Union's Horizon 2020 research and innovation programme under grant agreement No 757646 and A.P. additionally under grant agreement No 813942.
N.~H.-T.~is funded in part by the Albert Einstein Center for Fundamental Physics at the University of Bern, and in part by the Swedish Research Council, project number 2021-06638.
A.J. and J.R have been supported in part by UK STFC grant ST/P000711/1 and ST/T000775/1.
J.R. is also supported in part by UK STFC DiRAC operational grants ST/S003762/1 and ST/W002701/1.

\bibliographystyle{JHEP}
\bibliography{pl2}

\end{document}